# Test Time Reduction Reusing Multiple Processors in a Network-on-Chip Based Architecture

Alexandre M. Amory[1], Marcelo Lubaszewski[123], Fernando G. Moraes[4], Edson I. Moreno[4]
[1] PPGC - II - UFRGS - Av. Bento Gonçalves, 9500, Porto Alegre, RS - Brazil
[2.] PPGEE - EE - UFRGS, Av. Osvaldo Aranha, 103, Porto Alegre, RS - Brazil
[3.] IMSE-CNM–Universidad de Sevilla, Avda. Reina Mercedes, s/n, Sevilla - Spain
[4] PPGCC - FACIN - PUCRS - Av. Ipiranga, 6681, Porto Alegre, RS - Brazil
**amamory@inf.ufrgs.br, luba@{eletro.ufrgs.br,imse.cnm.es}, {moraes, emoreno}@inf.pucrs.br**

## Abstract

*The increasing complexity and the short life cycles of embedded systems are pushing the current system-on-chip designs towards a rapid increasing on the number of programmable processing units, while decreasing the gate count for custom logic. Considering this trend, this work proposes a test planning method capable of reusing available processors as test sources and sinks, and the on-chip network as the test access mechanism. Experimental results are based on ITC'02 benchmarks and on two open core processors compliant with MIPS and SPARC instruction set. The results show that the cooperative use of both the on-chip network and the embedded processors can increase the test parallelism and reduce the test time without additional cost in area and pins.*

## 1. Introduction

Recently, networks-on-chip (NoCs) emerged as an alternative communication architecture for complex system-on-chip (SoC) [3]. In addition, there is a rapid increasing on the number of programmable processing units in current designs. Considering this trend and the complexity for testing such systems, the reuse of both, the processor and the on-chip network for test, seems a promising approach to reduce test costs [1]. Software-based test of IP cores, i.e., the test based on the reuse of an embedded processor, presents several advantages, such as cheaper external test equipment, less test pins and minimal area overhead [4][5].

The reuse of embedded processors for test is not new, but no previous approach evaluated systems that support multiple processors on a parallel architecture. Huang *et al.* [4] evaluated the reuse of a MIPS processor on a bus-based architecture, while Hwang and Abraham [5] evaluated the reuse of an ARM processor on a Wishbone bus architecture. Amory *et al.* [2] developed a CAD tool, which helps the designer to integrate cores on a bus-based SoC and to generate test programs. The results were validated implementing the system in FPGA.

This paper proposes a new software-based test planning approach for NoC-based systems, which reuses the embedded processors and the NoC for testing.

## 2. Test Using NOCs

The process of using the NoC to carry the test patterns can be divided in three main steps. The first step corresponds to the *characterization of the NoC* in terms of time and power consumption. The performance metrics of a NoC router can be divided in two parts: the routing latency and the flow control latency. The *routing latency* is the intra-router time required to create a connection through the router, while the *flow control latency* is defined as the inter-router time required to send flits in the channels. In addition to the characterization of performance metrics, the designer also has to characterize the *power consumption* to send the test packets, which is identical to the power consumption of the NoC in functional mode. In our experiments, the power consumption has been measured as the mean power consumption to send packets of random size and random payload. This value is added to each router the packet passes through. Besides performance and power characterization, the designer also has to feed the tool with the topology, the routing algorithm, the number of routers and the flit width. Currently the tool supports NoCs based on grid topology using XY routing algorithm.

The second set of information the designer has to provide the tool with concerns the system. This set of information consists of the position of each core (including the processors reused for test), and the number and position of the IO ports that can be connected to the external tester. The position of the CUTs, processors and IO ports determine the order and priority of the test. The cores closer to IO ports or processors are tested first.

The second step comprises the *characterization of the processors* reused for test. A processor can be used during test in different ways. It can run a test program that reads the compressed test data from a memory, decompresses it and sends it to the core under test (CUT), or it can work as a test pattern generator emulating a pseudo-random BIST logic. Each of these test applications has to be modeled differently. Currently, we are modeling the BIST application, but in the near future we will also support decompression. In terms of *time*, the BIST application consumes time to generate the BIST pattern and to send it to the CUT. In terms of *power consumption*, the BIST application consumes power to generate patterns and to send it to the CUT. The test application has to be characterized in terms of time, memory requirements and power to each processor in the system reused for test. This step is necessary because the processors may have different instruction-sets, times to run the test application and power consumptions. Once this characterization is done, the results can be inserted into the proposed tool. In this work we modeled the Leon and Plasma processors, http://www.gaisler.com/ and http://www.opencores.org/, which are synthesizable VHDL models of processors compliant, respectively, with SPARC V8 and MIPS-I.

In addition to the characterization presented before, it is very important to take into account the test of the processor itself in the overall reuse strategy. Thus, the designer should provide the tool with the number of test patterns necessary to test each processor. A processor is reused for test just after it has been successfully tested. Complex processors require a large number of patterns to be tested, and may be reused for test few times, not contributing to reduce the global test time.

The last step comprises the *CUTs characterization*. We assume a core-based design methodology where the cores are acquired from the core providers to be integrated into the system chip. The core provider is responsible for the core test knowledge transfer. We assume this information is available with the core description since the proposed tool uses it for the global test planning.

## 3. Experimental Results

Experimental results for the proposed reuse approach are presented for three examples based on the benchmarks d695, p22810, and p93791 of the ITC'02 SoC Test Benchmarks set [6]. To system, cores representing the Leon and Plasma processors are added. For d695 system, six processor cores are added, whereas for p22810 and p93791 benchmarks, eight cores are added. The total number of cores of the new systems is 16, 36, and 40, respectively. The network dimensions for each system are, respectively, 4x4, 5x6 and 5x5.

Figure 1 presents the test time results for systems d695_Leon, p22810_Leon, and p93791_Leon, when different numbers of Leon processors are reused for test and two external interfaces (input and output) are used. Experiments with and without power constraints are presented for each system. This constraint is defined as a percentage of the sum of all cores power consumption. Thus, for example, a power limit of 50% indicates that the power limit corresponds to half of the sum of all cores power consumption in test mode. Notice that in a real case, the designer can define any power limit.

The results presented in Figure 1 demonstrate that increasing the number of processors reused for test reduces the test time compared to the test without processor reuse. One can observe that even smaller systems like d695_leon can take advantage of the extra test interface, with test time reduction of 28%. For larger systems such as p93791_leon, the gain in test time can be as high as 44%. Despite of this, imposing power constraints the test reduction reaches up to 37%, since there is more parallelism and more processors running the test programs.

For the p22810_leon system, we get some test time reduction, but it is not regular because of the greedy behavior of the scheduling algorithm. The greedy behavior of the presented algorithm forces it to select the first test interface available. This can increase the test time because we assume the processor takes 10 clock cycles to generate a test pattern, while the external tester takes zero clock cycles. Thus, if a processor is available in a given instant and an external tester is available a few instants later, the resource used will be the processor, since it was available before. However, the external tester should be used because it is faster than the processor.

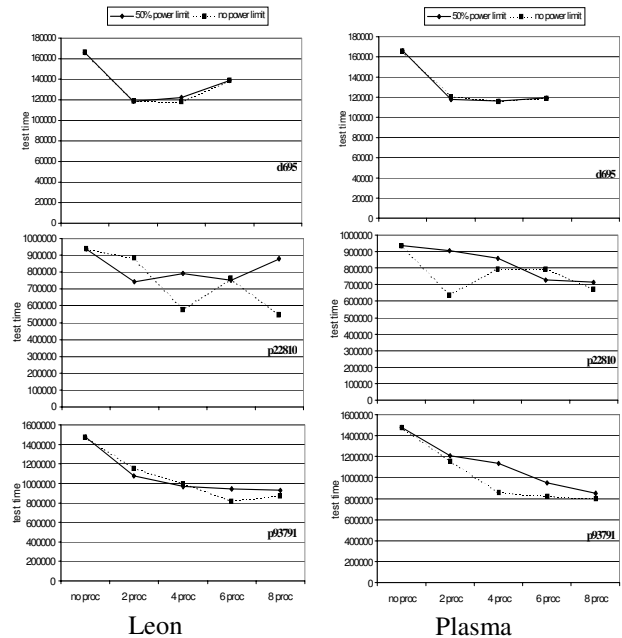

**Figure 1 - Test times.**

## 4. Final Remarks

This work evaluates the impact of reusing processors to test a NoC-based system. The processors available in the system are programmed with an application to test cores of the system. The results presented demonstrate that, increasing the number of test sources/sinks to explore the NoC parallelism, reduce the system test time without additional area and test pins.